\documentstyle[aps,manuscript]{revtex}

\begin{document}
\title {Mass and Isotope Dependence of Limiting Temperatures for Hot Nuclei \footnote 
{E-Mail: lizwux@iris.ciae.ac.cn}}
\author{Zhuxia Li$^{1,2,3}$, Min Liu$^{1}$ }
\address{ 1) China Institute of Atomic Energy, P. O. Box 275(18), Beijing\\
102413\\
2) Nuclear Theory Center of National Laboratory of Heavy Ion Accelerator,\\
Lanzhou 730000\\
3) Institute of Theoretical Physics, Chinese Academic of Science,
Beijing
100080\\
}
\maketitle

\begin{abstract}
The mass and isotope dependence of limiting temperatures for hot
nuclei are investigated. The predicted mass dependence of limiting
temperatures is in good agreement with data derived from the
caloric curve data. The predicted isotope distribution of limiting
temperatures appears to be a parabolic shape and its centroid is
not located at the isotope on the $\beta$-stability line(T=0) but
at neutron-rich side. Our study shows that the mass and isotope
dependence of limiting temperatures depend on the interaction and
the form of surface tension and its isopin dependence sensitively.
\end{abstract}
\begin{center}
{\bf 1. INTRODUCTION}
\end{center}
It is well known that for nuclear matter, liquid and gas phases
can coexist at temperature lower than $T_{c}$ and only gas phase
can exist at or above $T_{c}$. While for a hot nucleus surrounding
by vapor, due to the Coulomb instabity\cite{Ja83,Bo84,Bo86,Le85},
it may not be stable above a temperature which is much lower than
the critical temperature $T_{c}$ of infinite nuclear matter, this
temperature is called the limiting temperature $T_{lim}$. Limiting
temperatures for nuclei along the $\beta$-stability line were
studied by using Skyrme interaction
\cite{Ja83,Bo84,Be89,So91,Zh96,Liu03} and Furnstahl, Serot and
Tang (FST) model\cite{Zh99}. Their studies showed that the
limiting temperature was sensitive to the effective
nucleon-nucleon interactions. In ref.\cite{Na02a,Na02b} the mass
dependence of limiting temperatures for hot nuclei was derived by
analyzing the existing caloric data. The establishment of the beam
of rare isotopes will provide us the opportunities to study the
heavy ion collisions with large isospin asymmetry. The
measurements of caloric curves(or other possible observables) can
possibly explore the limiting temperatures for hot nuclear systems
with large isospin asymmetry. The limiting temperatures for nuclei
away from the $\beta$-stability line should depend on the isospin
dependent part of the Equation of State(EOS). The newly developed
Skyrme forces such as the $SLy$ series are designed to study the
properties of nuclei away from the $\beta$-stability line in
addition to nuclei along the  $\beta$-stability line. It has been
shown that these modern Skyrme forces can describe the properties
of the nuclei away from $\beta$-stability line much better than
the old Skyrme forces\cite{Ch97}. Therefore it seems to us to be
worthwhile to apply the modern parametrization of Skyrme force
such as the $SLy$ series for studying the limiting temperatures
for nuclei both along and away from the $\beta$- stability line.

Additionally, the surface tension is another important factor
affecting the limiting temperature. The surface tension represents
a work per unit area needed to create the surface, the work is
required because the nucleons at the surface are less bounded.
Therefore the surface tension is related to the EOS of nuclear
matter. The symmetry energy term in the EOS should also contribute to
the surface tension.  However, the relationship is not available and
it is treated independently of the equation of state except the
fact that the surface tension vanishes as $T \ge T_{c}$. In this
case, the form of the temperature dependence of the surface tension
is not unique in literatures and little attention has been paid to
the influence of the different forms of the surface tension on the
limiting temperature. The isospin dependence of the
surface tension is even unclear. Therefore it is requisite to investigate the
effect of the different forms of the surface tension, including with
and without the isospin dependence(surface symmetry term), on limiting temperature. By
comparing the calculation results with the experimental data we
may obtain the information of the surface tension and better
parametrization of Skyrme interaction.

In this work we follow the approach of Ref.\cite{Ja89a,Ja89b}
to study the mass and isotope dependence of limiting
temperatures by using the Skyrme effective force of $SLy7$ in addition to SIII and
$SkM^{*}$.  The effect of
different forms of the surface tension on the limiting temperature
will also be investigated. In Sec. II, The equation of state by using $Sly7$
is studied.   In Sec.III we
give the coexistence equations and the calculation results of the
mass and isotope dependence of limiting temperatures of hot
nuclei. Finally, a summary and discussions is given in Sec.IV.

\begin{center}
{\bf 2. THE EQUATION OF STATE}
\end{center}
The single-particle energy reads
\begin{equation}
\varepsilon_{q}=\frac{\hbar^{2}k^{2}}{2m_{q}^{*}}+u_{q}+\varepsilon _{Coul}\delta _{q,p},
\end{equation}
where $m_{q}^{*}$ and $u_{q}$ is the effective mass and the
single-particle potential energy of species q, respectively. The third term is
the Coulomb energy. The chemical potential $\mu _{q}$ of species q
for nuclear system at temperature $T$ can be written as
\begin{equation}
\mu _{q}(T,\rho _{n},\rho _{p})=u_{q}(\rho _{n},\rho
_{p})+\varepsilon _{Coul}\delta _{q,p}+T\ln \frac{\lambda _{T}^{3}
}{g_{s}}\rho _{q}+T\sum _{n=1}^{\infty}\frac{n+1}{n}b_{n}(\frac{
\lambda _{T}^{3}}{g_{s}}\rho _{q})^{n}.
\end{equation}
Here $\lambda_{T}$ is the effective
thermal wavelength of the nucleon, which reads
\begin{equation}
\lambda_{T}=(\frac{2\pi\hbar^{2}}{m_{q}^{*}T})^{\frac{1}{2}},
\end{equation}
and $b_{n}$'s are the coefficients of the virial series for ideal Fermi gas.
For the most general form of a Skyrme-type interaction, $u_{q}$ can be written as
\begin{eqnarray}
&& u_{q}=A_{1}\rho +A_{2}\rho _{q}+A_{3}\tau +A_{4}\tau
_{q}+A_{5}(\alpha +2)\rho ^{\alpha +1}
                                     \nonumber\\
&& +A_{6}\alpha \rho ^{\alpha -1}(\rho _{n}^{2}+\rho
_{p}^{2})+2A_{6}\rho ^{\alpha }\rho _{q}.
\end{eqnarray}
$\rho$ and $\tau$ are the nuclear density and kinetic energy density,respectively,
 and $\rho_{q}$ and $\tau _{q}$ are the density and the kinetic energy density of species $q$,respectively.
 $A_{1}$ - $A_{6}$ are
\begin{eqnarray}
 && A_{1}=t_{0}(1+\frac{1}{2}x_{0}),
                            \nonumber\\
 && A_{2}=-t_{0}(\frac{1}{2}+x_{0}),
                             \nonumber\\
 && A_{3}=\frac{1}{4}[t_{1}(1+\frac{1}{2}x_{1})+t_{2}(1+\frac{1}{2}x_{2})],
                                                                  \nonumber\\
 &&A_{4}=\frac{1}{4}[t_{2}(\frac{1}{2}+x_{2})-t_{1}(\frac{1}{2}+x_{1})],
                                                               \nonumber\\
 &&A_{5}=\frac{1}{12}t_{3}(1+\frac{1}{2} x_{3}),
                                       \nonumber\\
 &&A_{6}=-\frac{1}{12}t_{3}(\frac{1}{2}+x_{3}),
\end{eqnarray}
where $t_{0},t_{1},t_{2},t_{3}, x_{0},x_{1},x_{2},x_{3}$ are the parameters of the Skyrme interaction.
The effective mass $m_{q}^{*}$  reads
\begin{equation}
m_{q}^{*}=m[1+\frac{2m}{\hbar^{2}}(A_{3}\rho+\frac{1}{2}A_{4}(1\pm
y)\rho)]^{-1}.
\end{equation}

Introducing
$y=\frac{\rho_{n}-\rho_{p}}{\rho_{n}+\rho_{p}}$
we obtain
\begin{eqnarray}
&& \mu_{q}(T,\rho,y)=u_{q}(\rho,y)+\varepsilon_{Coul}(\rho)\delta_{q,p}
+Tln\frac{\lambda_{T}^{3}}{g_{s,I}}\rho+Tln(1 \pm y) \nonumber\\
&&+T\sum_{n=1}^{\infty}\frac{n+1}{n}b_{n}(1 \pm y)^{n}
(\frac{\lambda_{T}^{3}}{g_{s,I}}\rho)^{n}.
\end{eqnarray}
and
\begin{eqnarray}
 && \tau_{q}=\frac{g_{s}}{(2\pi )^{3}} \int d^{3}k n_{q}(k)k^{2}
                                                            \nonumber\\
 &&=\frac{g_{s}}{(2\pi )^{3}} \int d^{3}k [1+exp((\varepsilon_{q}-\mu_{q})/T)]^{-1}k^{2}
                                                                               \nonumber\\
 &&=\frac{3}{2}T\frac{m_{q}^{*}}{\hbar^{2}}\rho\sum_{n=0}^{\infty}b_{n}
 (1\pm y)^{n+1}(\frac{ \lambda_{T}^{3}}{g_{s,I}}\rho)^{n},
\end{eqnarray}
where the symbol $"+"$ stands for neutrons and $"-"$ for protons.
Concerning the expansion in the degree of the degeneracy ($\frac{
\lambda_{T}^{3}}{g_{s,I}}\rho$) in $\mu _{q}$ and $\tau_{q}$, in
this work, $n \ge 7$ term  are neglected, i.e. one more term
than in Refs.\cite{Ja89a,Ja89b} is included. We find that for
temperatures and densities relevant to this work(say,T $\sim
6MeV$ and $\rho \sim 0.14 fm^{-3}$) at least 6 terms are needed
in order to obtain accurate results. $g_{s}$ is the spin
degeneracy ($g_{s}=2$) while $g_{s,I}$ is the spin-isospin
degeneracy ($g_{s,I}=4$).

Assuming that a hot nucleus is a spherical drop with a sharp edge
and uniform distribution of nucleons,the Coulomb energy
can be expressed as
\begin{equation}
\varepsilon _{Coul}=\frac{6}{5}\frac{Ze^{2}}{
R_{L}}-(\frac{3}{2\pi
})^{\frac{2}{3}}\frac{Z^{\frac{1}{3}}e^{2}}{R_{L}}.
\end{equation}
Here the Coulomb exchange term is included.  $R_{L}$ is the
radius of the nucleus and expressed as
\begin{equation}
 R_{L}=(\frac{3A}{4\pi \rho })^{\frac{1}{3}}.
\end{equation}
The symbols $Z$ and $A$ are the number of protons and nucleons in
the liquid drop, respectively.

The pressure can be obtained through the Gibbs-Duhem relation:
\begin{equation}
\frac {\partial\widetilde{P}} {\partial\rho}
=\rho_{n}\frac{\partial \mu _{n}}{\partial \rho
}+\rho_{p}\frac{\partial \mu _{p}}{\partial \rho }.
\end{equation}
Thus,
\begin{eqnarray}
\widetilde{P}(T,\rho,y)=\frac{1}{2}\sum _{q=p,n}(1\pm y)\int
\limits_{0}^{\rho }\rho\frac{\partial \mu _{q}}{\partial \rho
}\,d\rho,
\end{eqnarray}
where symbol $"+"$ stands for neutrons and $"-"$ for protons.

The surface tension term has to be introduced in the pressure for a
finite system. For the nuclei along the $\beta$-stability line we
adopt two different forms of the surface tension. The first one is:

\begin{equation}
\hspace{1cm}\gamma(T)=1.14MeV\cdot
fm^{-2}(1+\frac{3}{2}\frac{T}{T_{c}})(1-\frac{T}{T_{c}})^{\frac{3}{2}}
\end{equation}
suggested by Ref.\cite{Go84} and used by
Ref.\cite{So91,Zh96,Zh99,Ja89a,Ja89b} in studying the limiting temperature.
Here we call it Surf1.
The other one called Surf2 is:
\begin{equation}
\hspace{1cm}\gamma(T)=\gamma(0)
[(T_{c}^{2}-T^{2})/(T_{c}^{2}+T^{2})]^{\frac{5}{4}}
\end{equation}
\begin{equation}
\gamma(0)\approx 18MeV/4\pi r_{0}^{2},\hspace{1cm}r_{0}=1.12fm,
\end{equation}
which was suggested by \cite{Bo85}. This expression of the surface
tension is widely used in SMM calculations for studying
multifragmentation process in heavy ion collisions. It was obtained
as a parametrization of the calculation of thermodynamic
properties of the interface between liquid and gas of symmetrical
nuclear matter which were performed with Thomas-Fermi and H-F
method by using Skyrme force. $T_{c}$ is the critical temperature for infinite nuclear matter.
The pressure caused by the surface
tension ($P_{Surf}$) is given by:
\begin{equation}
P_{Surf}=-2\gamma(T)/R_{L}.
\end{equation}
Thus, the total pressure of the liquid drop ($P$) is written as
\begin{equation}
P(T,\rho,y)=\widetilde{P}(T,\rho,y)+P_{Surf}.
\end{equation}
Fig.1 shows the isothermal curves of the chemical potential
calculated with SLy7 for: a)protons with bulk part only,
b)neutrons, c)protons with bulk part adding the Coulomb term of a
hot $^{208}Pb$ nucleus. Fig.2 shows the isothermal curves of the
pressure calculated with SLy7, in which the Coulomb and the
surface tension term are for a hot $^{208}Pb$ nucleus. Fig.3 shows
the pressure as a function of density at T=5MeV calculated with
$SLy7$, $SkM^*$ and $SIII$ in which the Coulomb term and the
surface tension term are for a hot $^{208}Pb$ nucleus. There are
three sets of curves, namely, the bulk, the
bulk+Coulomb($^{208}Pb$) and the bulk+Coulomb($^{208}Pb$)+surface
tension($^{208}Pb$). In each set, there are three curves
corresponding to $SLy7$, $SkM^*$ and $SIII$, crossing at density
around $\rho \sim 0.13fm^{-3}$. One sees a considerable negative
pressure provided by the surface tension through comparing the
sets with and without surface tension. Furthermore, one can find
from the figure that the curves for $SLy7$ and $SkM^*$ are very
close at the densities relevant to this study, while the curve for
SIII is far away from them except at the crossing point. To show
the effect of the different forms of the surface tension, in Fig.4 we plot
the pressure at T=5 MeV for $^{208}Pb$ calculated with the surface
tension of Surf1, Surf2, and without a surface tension. One can
find that the Surf1 produces a stronger negative pressure than the
Surf2 does, which should have effect
on the limiting temperature especially for light nuclear systems.\\
\begin{center}
{\bf III. MASS AND ISOTOPE DEPENDENCE OF LIMITING TEMPERATURES }
\end{center}
We adopt the the same model used in refs.\cite{Ja89a,Ja89b} to
study the mass and isotope dependence of limiting temperatures for
hot nuclei. The model treats the hot nucleus as a uniformly charged
drop of nuclear liquid with sharp edge at a given temperature,
which is in thermal equilibrium with the surrounding vapor. The
equilibrium between the droplet and the vapor surrounding both
thermal mechanical and chemical leads to a set of two-phases
coexistence equation:
\begin{equation}
\mu_{p}(T,\rho_{L},y_{L})=\mu_{p}(T,\rho_{V},y_{V}),
\end{equation}
\begin{equation}
\mu_{n}(T,\rho_{L},y_{L})=\mu_{n}(T,\rho_{V},y_{V}),
\end{equation}
\begin{equation}
\widetilde{P}(T,\rho_{L},y_{L})+P_{Surf}(T,\rho_{L})=P(T,\rho_{V},y_{V}).
\end{equation}
The subscript $L$ refers to the liquid phase and $V$ to the vapor
phase. For simplification, the Coulomb interaction in vapor is
screened in the calculation of the pressure $P(T,\rho_{V},y_{V})$ and
the chemical potential of protons $\mu_{p}(T,\rho_{V},y_{V})$.
These three coexistence equations with three variables can be
solved directly to get the coexistence point of the liquid drop and
the surrounding vapor. By finding the upper
boundary of temperature that the coexistence equations have
solution one obtains the limiting temperature.

\begin{center}
{\bf 1. Mass Dependence of Limiting Temperatures}
\end{center}
Now let us study the mass dependence of limiting temperatures
of nuclei along the $\beta$-stability line:
\begin{equation}
Z=0.5A-0.3\times 10^{-2}A^{\frac{5}{3}}.
\end{equation}
Fig.5 - Fig.7 show the mass dependence of limiting temperatures
calculated with surface tension Surf1 and Surf2 and Skyrme
interactions $SLy7$, $SkM^*$, $SIII$, respectively. The data of
\cite{Na02a,Na02b} are also plotted in the figures. From these 3
figures one can see that limiting temperatures decrease with
the system size, which is independent of the force used.
Furthermore, one finds that the limiting temperatures calculated
with $SLy7$ and $SkM^{*}$ are very close and the largest
difference is about or less 0.1 MeV while the limiting
temperatures calculated with SIII are much higher. This is because
the EOS corresponding to $SLy7$ and $SkM^{*}$ is of soft one while
that for $SIII$ is of stiff one(see Fig.3). Concerning the surface
tension, one can see that the limiting temperatures calculated
with Surf2 are higher than that with Surf1.  We find, in general, the
calculation results with surface tension Surf2 and Skyrme force
$SLy7$ agree with experimental data better than other combinations
of surface tension forms and interaction parameter sets. This
agreement may illuminate that the surface tension may be better
described by Surf2 rather than Surf1 at $T<T_{c}$. Therefore. in
the following calculations we adopt $SLy7$ and Surf2. In Table.1
we list the values of limiting temperatures for the nuclei along
the $\beta$-stability line with mass number ranging from 20 to
250.

\begin{center}
{\bf 2. Isotope Distribution of Limiting Temperatures}
\end{center}

Fig. 8 shows the isotope distribution of limiting temperatures for
nuclei: a)$O$, b)$Ca$, and c)$Zr$ calculated with Skyrme
interaction $SLy7$ and surface tension Surf2, respectively. One
can see from these figures that the isotope distribution of
limiting temperatures appears to be a parabolic shape. The
centroid of the parabolic curve is not located at the isotope on
$\beta$-stability line(T=0) but incline to the neutron-rich side.
The shift of the centroid of the isotope distribution of limiting
temperatures to the neutron-rich side seems to be in consistent
with\cite{Na95} in which it was mentioned that there was evident
that neutron rich nuclei had higher limiting temperatures. This
effect is due to the fact that the Coulomb term does not depend on
the temperature explicitly while the isospin dependent part of the
bulk term does depend on the temperature explicitly, which leads
the compensation effect of the Coulomb energy and the symmetry
energy with respect to the isospin degree to be different for hot
nuclei and cold nuclei.

It has been observed that the isospin fractionation distillation effect
on multifragmentation process in
neutron-rich heavy ion collisions\cite{Xu00}. We expect that this
effect should also be shown up in the
equilibrium value of the isospin asymmetry of vapor phase at the
limiting temperature. In Fig.9 we show the isotope
distribution of the equilibrium values of the isospin asymmetry of
vapor at the limiting temperature, $y_{v}^{lim}$, for $O$, $Ca$, and $Zr$, respectively. One
can see from the figure that the equilibrium value of $y_{v}^{lim}$
increases with the increase of
the number of neutrons and finally a saturation value might be
reached. The isospin distillation effect leads vapor to be even
neutron-rich for neutron-rich isotopes and to be even neutron-lack
for neutron-lack isotopes which can be seen in the figure clearly.
The vanish point of the equilibrium value of $y_{v}^{lim}$,
at which the isospin asymmetry of vapor changes from
negative to positive value, is shifted a little to the
neutron-rich side. The vanish point should be sensitive to the
isospin dependent part of both the interaction and the surface
tension.

As is already mentioned in the introduction that the
$SLy$ series of Skryme interactions is designed to describe the properties of
isospin asymmetric nuclei, its
isospin dependent part is rather different from
$SkM^{*}$\cite{Ch97}. We expect that the isospin distribution of
limiting temperatures calculated with $SLy7$ should be different
from that with $SkM^{*}$ though the mass distributions of
limiting temperatures calculated with $SLy7$ and $SkM^{*}$ are very
close.  Our calculation results of the isotope distribution
of limiting temperature of $Zr$ shown in Fig.10 does demonstrate that the isotope
distribution of limiting temperatures depends on the isospin dependent part of
interaction sensitively.
Limiting temperatures calculated with  $SLy7$ are higher than those with
$SkM^{*}$ and the difference between them increases as the isospin asymmetry
$|N-Z|/A$ increases. It was demonstrated that the $SLy$ series of
interactions had a  better isotope  properties than old Skyrme
interactions\cite{Ch97}, so the isotope distribution predicted
with $SLy7$ may be more proper.

  As pointed out at Sec.I, the surface tension is a work per surface area needed to create
the surface.  The symmetry energy term in EOS should also contribute to the surface tension and
therefore a symmetry-surface tension term should be introduced in the surface tension.
Here we adopt a surface tension including a symmetry-surface term:
 \begin{equation} \hspace{1cm}\gamma(T)=(\gamma(0)-a_{s}y^{2})
[(T_{c}^{2}-T^{2})/(T_{c}^{2}+T^{2})]^{\frac{5}{4}}.
\end{equation}
to study the effect of the isospin dependence of the surface tension. Here $a_{s}$ is
taken to be 28.5MeV\cite{Ch97,My66}. Fig.11 shows the isotope
distribution of limiting temperatures of hot nuclei $C$ and $O$
calculated with and without
symmetry-surface tension term taken into account, respectively. The
symmetry-surface tension term raises limiting temperatures of
neutron-rich isotopes but almost does not affect limiting
temperatures at the neutron-lack side. Therefore the information of
the symmetry-surface tension term can only be obtained by the measurement
of limiting temperatures of neutron-rich nuclei.

\begin{center}
{\bf IV. SUMMARY}
\end{center}
In summary, in this work we have studied the mass and isotope dependence of
limiting temperatures. The tendency of limiting temperatures
decreasing with the increasing of mass number for nuclei along the
$\beta$-stability line is in agreement with refs.\cite{So91,Zh96,Zh99}.
The influence of different Skyrme forces and different forms of
the surface tension on limiting temperatures of hot nuclei is
investigated. We find that the results calculated with newly developed Skyrme
force $SLy7$ and the surface tension of Surf2 are in good
agreement with data. The isotope distribution of limiting
temperatures appears to be a parabolic shape. The centroid of the
parabolic curve is not located at the isotope of $\beta$-stability line
but inclined to the neutron-rich side. The equilibrium value of
the isospin asymmetry of vapor phase at the limiting temperature
$y_{v}$ shows clearly the isospin fractionation distillation
effect. An isospin dependent surface tension is introduced in studying
the isotope distribution of limiting temperatures.
Our study shows that the neutron-rich side of the isotope distribution of limiting
temperatures depends on the isospin dependence of the
the surface tension sensitively.
\begin{center}
{\bf ACKNOWLEDGMENTS}
\end{center}
This work is Supported by National Natural Science Foundation of
China under Grant Nos. 10175093, 10235030, Science Foundation of
Chinese Nuclear Industry and Major State Basic Research
Development Program under Contract No. G20000774.
\newpage
\begin{center}
{\bf Figures Caption }
\end{center}

Fig.1 The isothermal curves of the chemical potential
calculated with SLy7 for: a)protons with bulk
part only, b)neutrons, c)protons with bulk part adding the Coulomb term
of a hot $^{208}Pb$ nucleus.

Fig.2 The isothermal curves of pressure. The bulk part is calculated with SLy7.
The Coulomb and the surface tension part are of
a hot $^{208}Pb$ nucleus.

Fig.3 The isothermal curve of pressure at T=5MeV  calculated
with $SLy7$, $SkM^*$, and $SIII$, respectively. The Coulomb and surface tension
term is corresponding to a hot $^{208}Pb$ nucleus.

Fig.4 The comparison of the pressure $P$ at $T=5MeV$ for
three cases, namely, without surface tension, with Surf1 and
Surf2. The surface tension term is corresponding to a hot $^{208}Pb$ nucleus.

Fig.5 The mass dependence of the
limiting temperatures calculated with Skyrme interaction $SLy7$ and
surface tension Surf1 and Surf2, respectively.

Fig.6 The same with Fig.6 but with $SKM^*$.

Fig.7 The same with Fig.6 but with $SIII$.

Fig.8 Isotope distributions of the limiting temperatures for $O$, $Ca$,
and $Zr$.

Fig.9 Isotope distributions of the equilibrium values of the
isospin asymmetry of vapor at limiting temperature for
$O$,$Ca$,$Zr$, respectively.

Fig.10 Isotope distribution of the limiting
temperatures for $Zr$ calculated with $SLy7$ and $SkM^{*}$,
respectively.

Fig.11, the isotope distribution of limiting temperatures of hot nuclei $C$ and $O$
calculated with and without symmetry-surface term taken into account, respectively.

Table.1 The values of limiting temperatures for nuclei along the $\beta$-stability line.

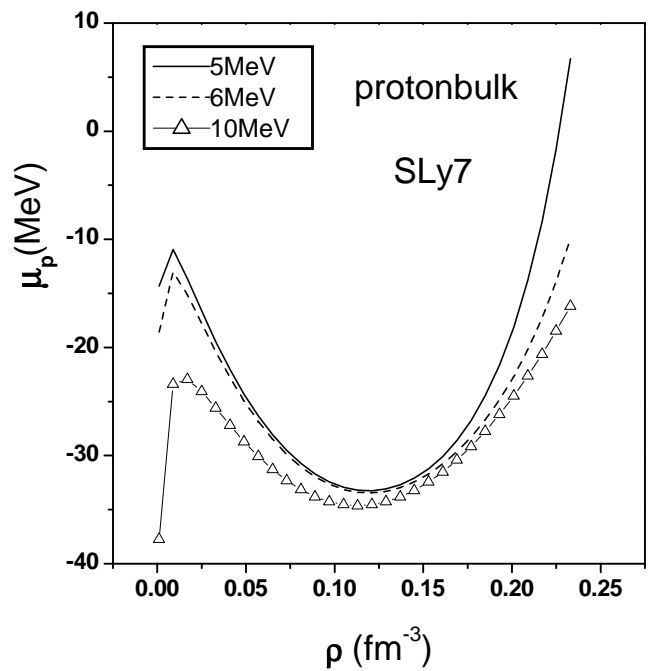
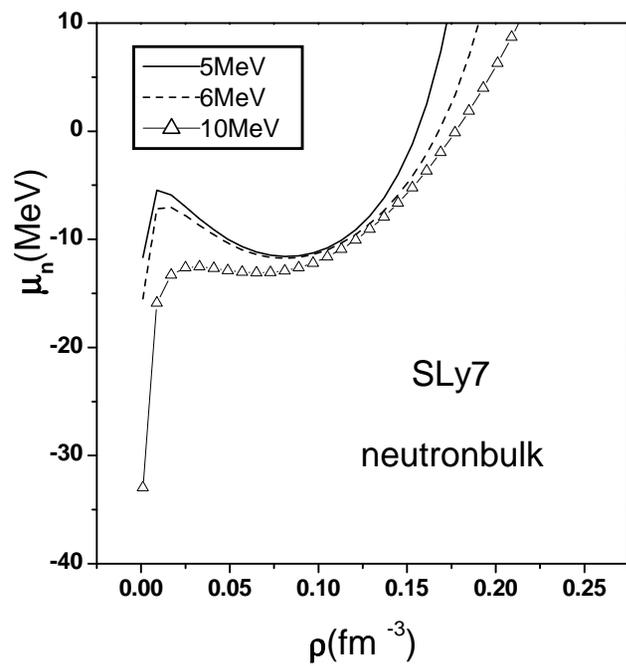
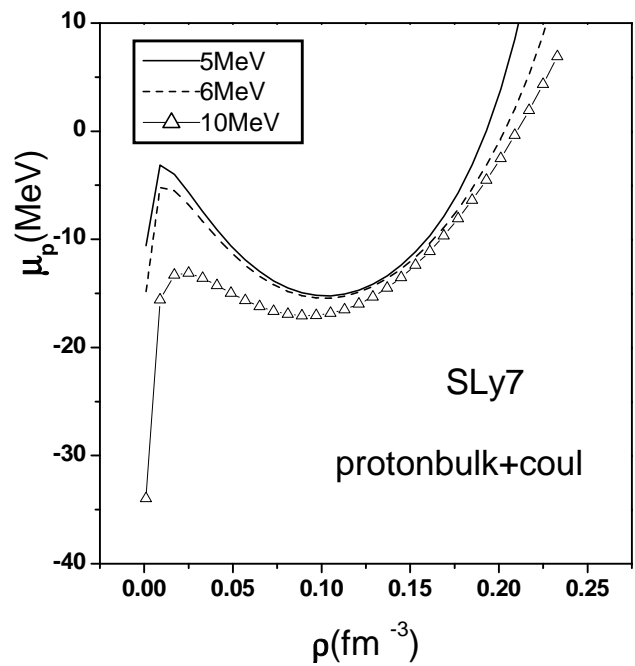
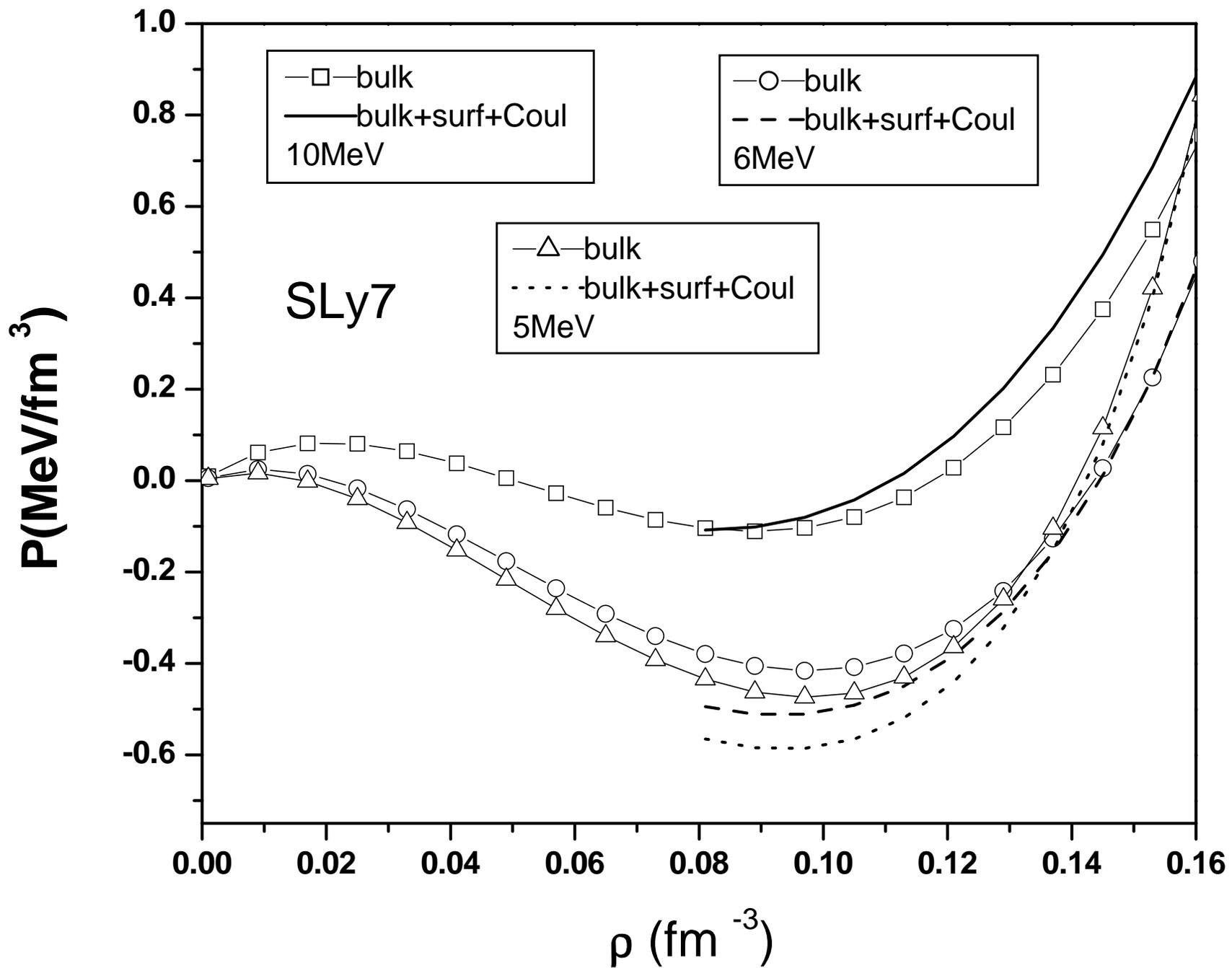
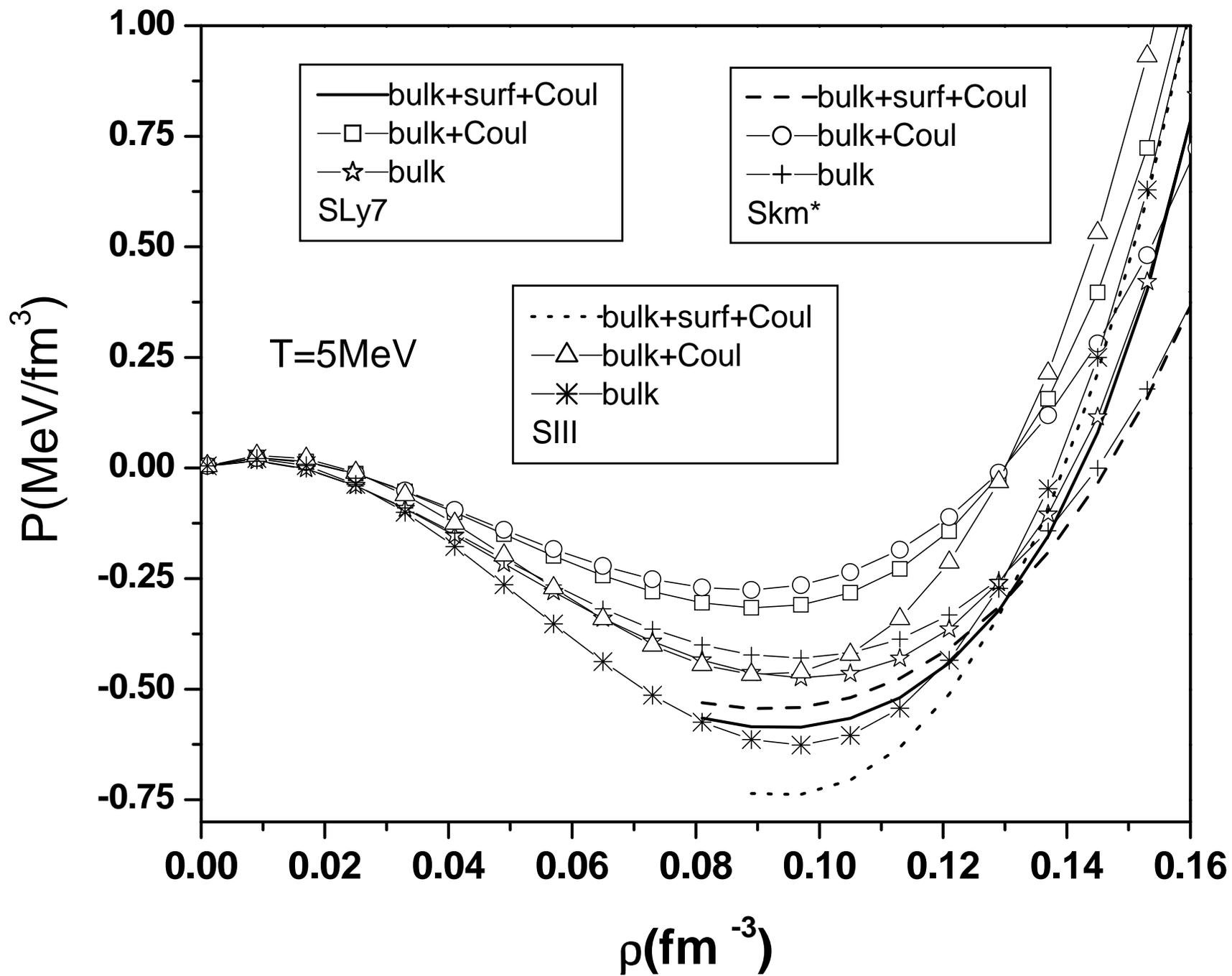
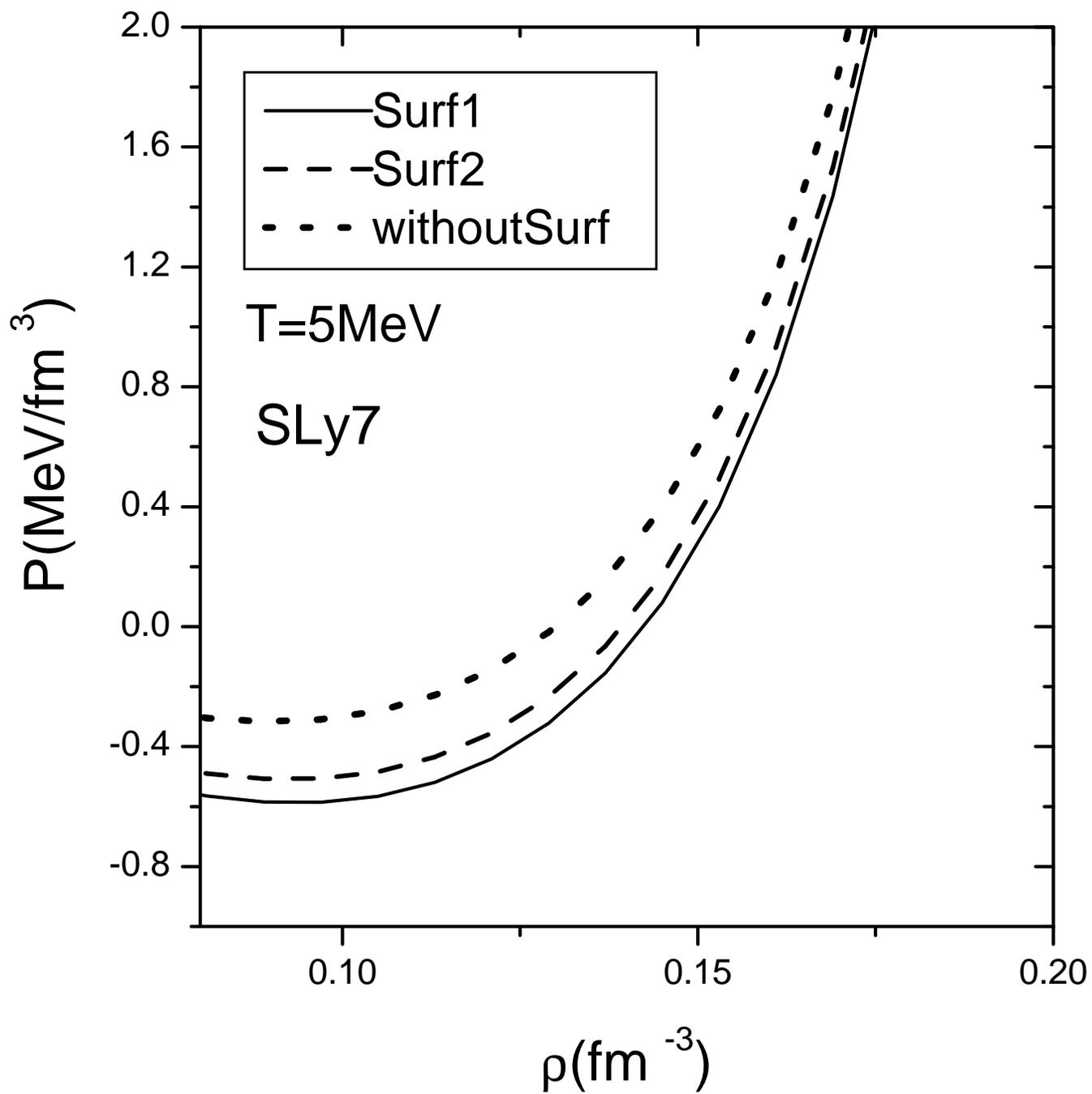
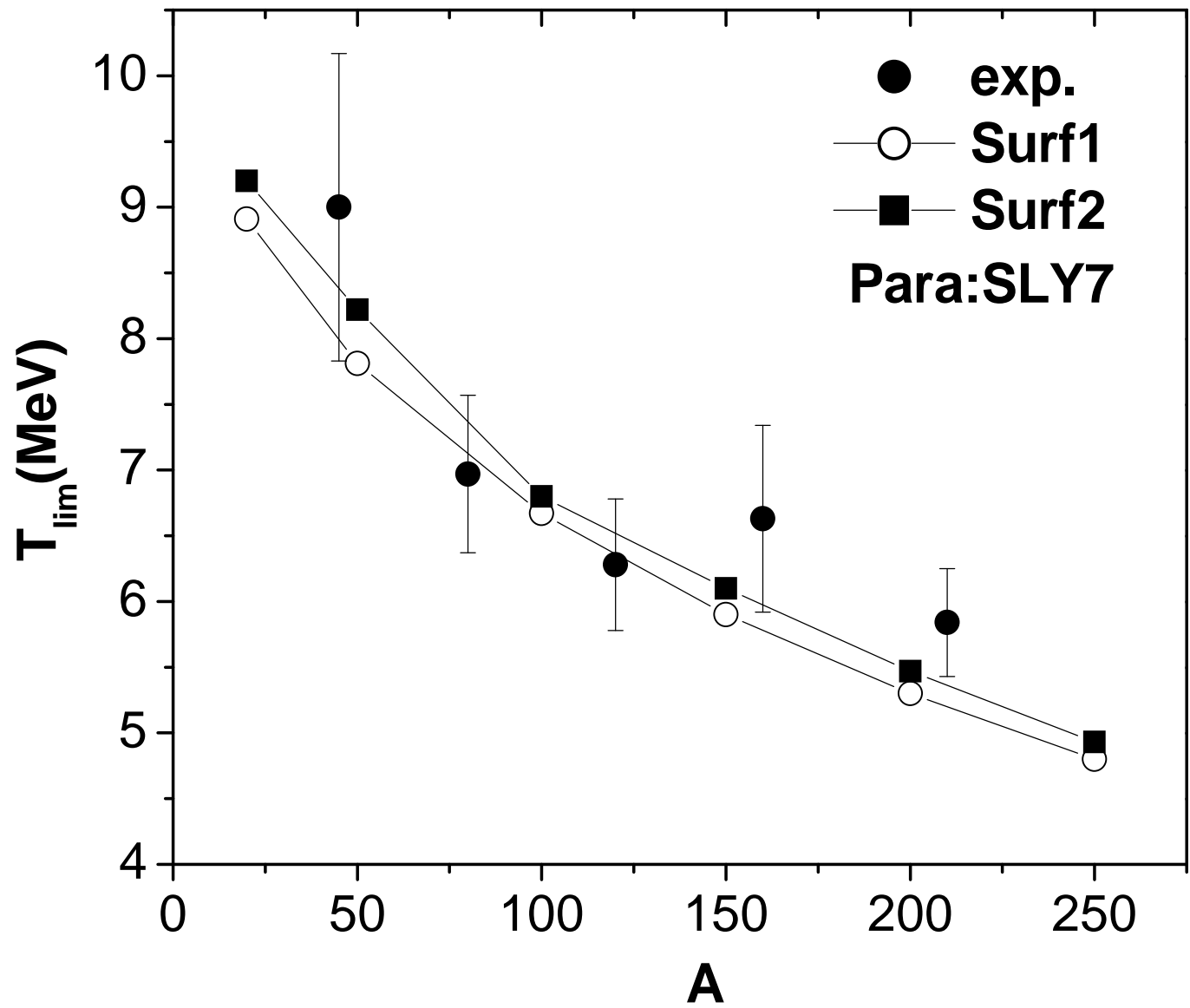
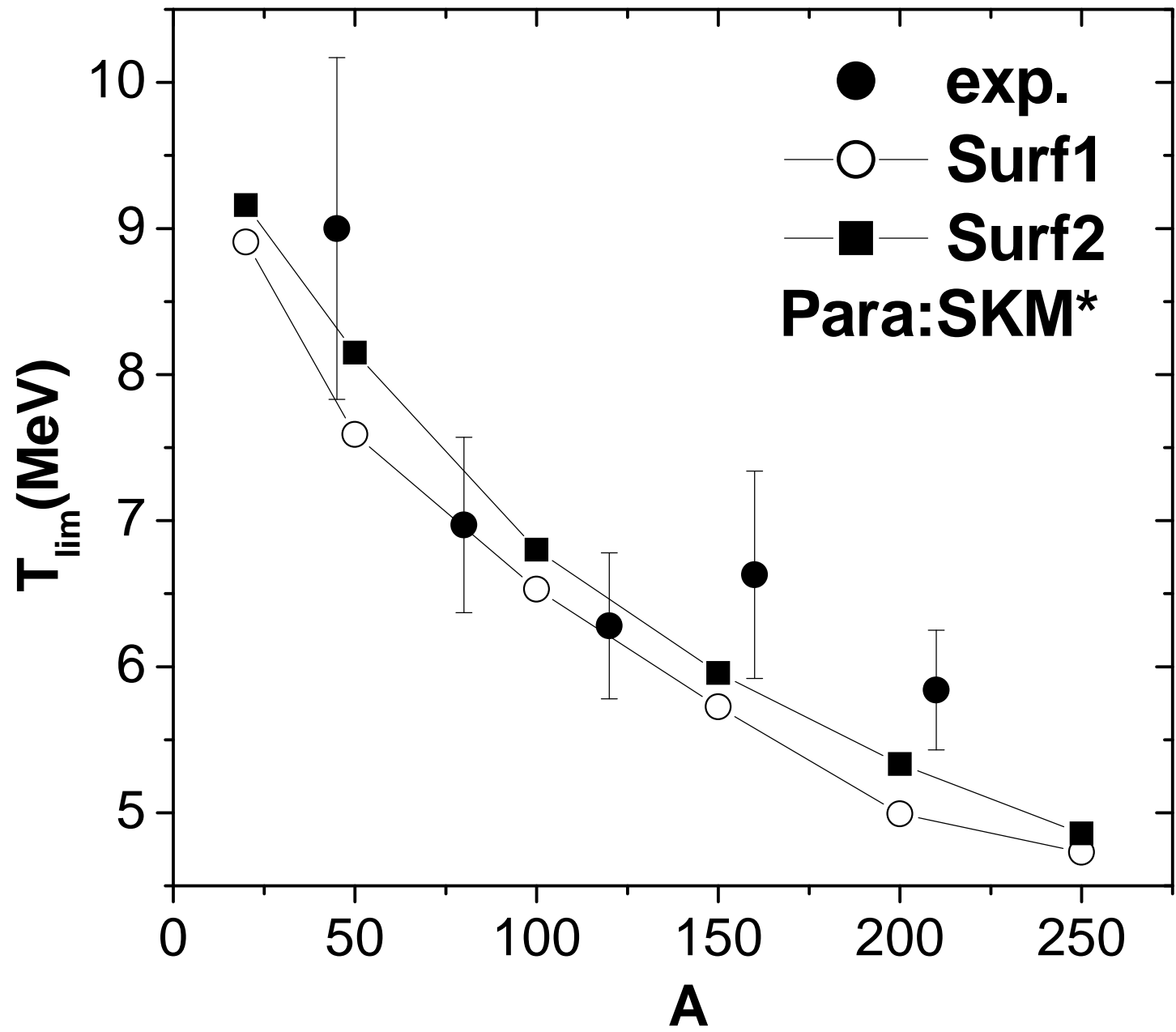
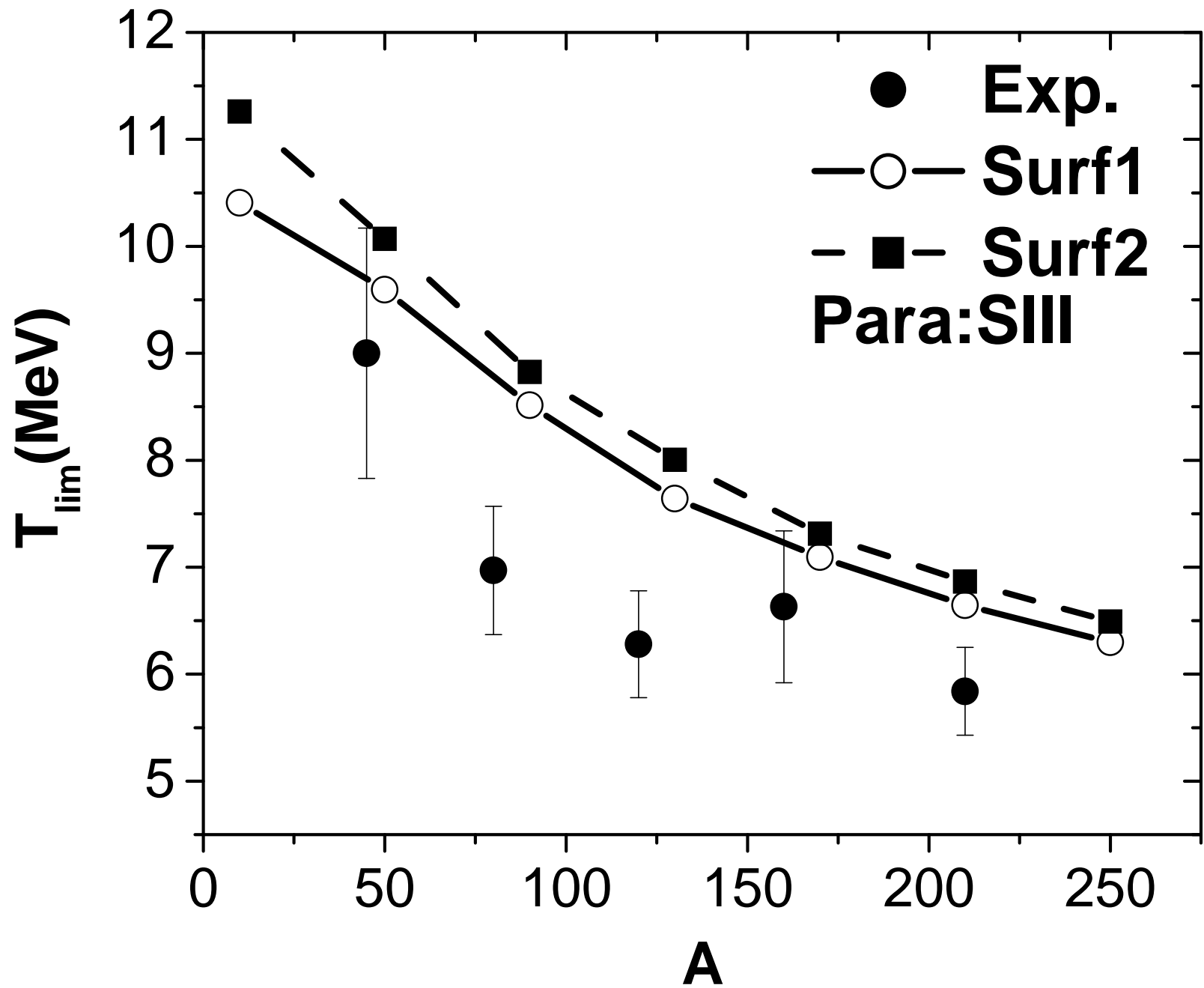
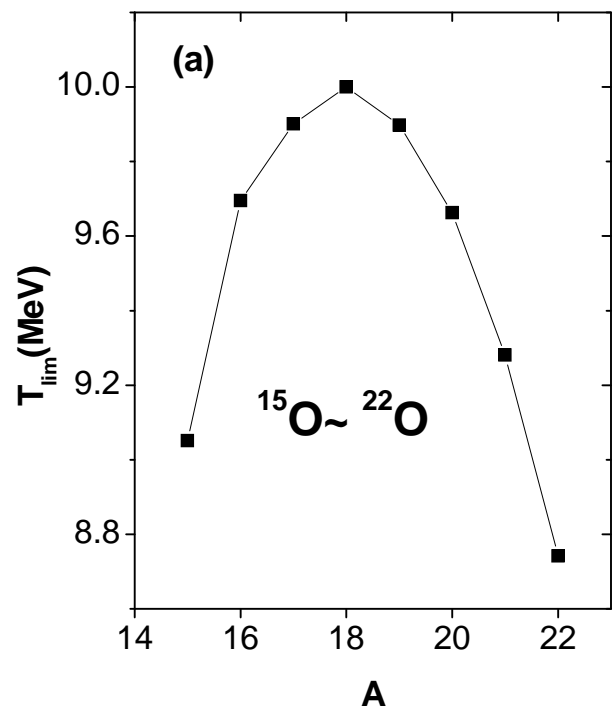
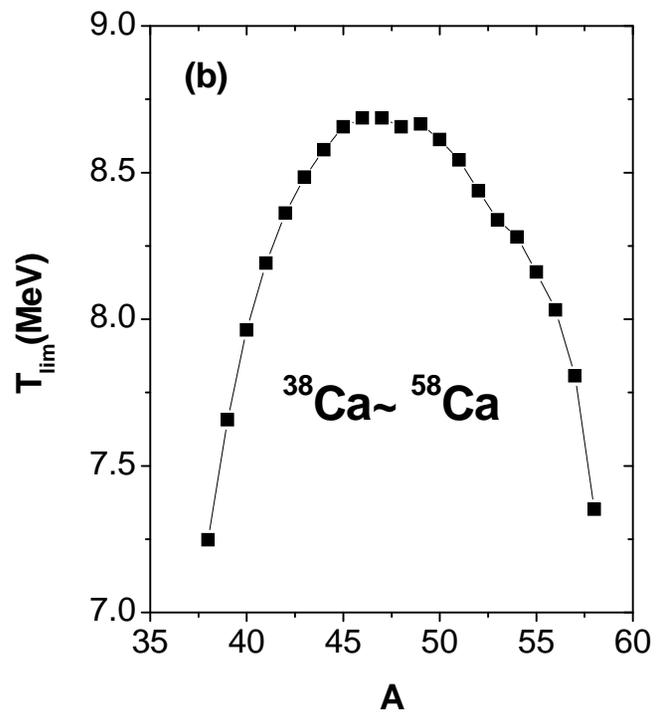
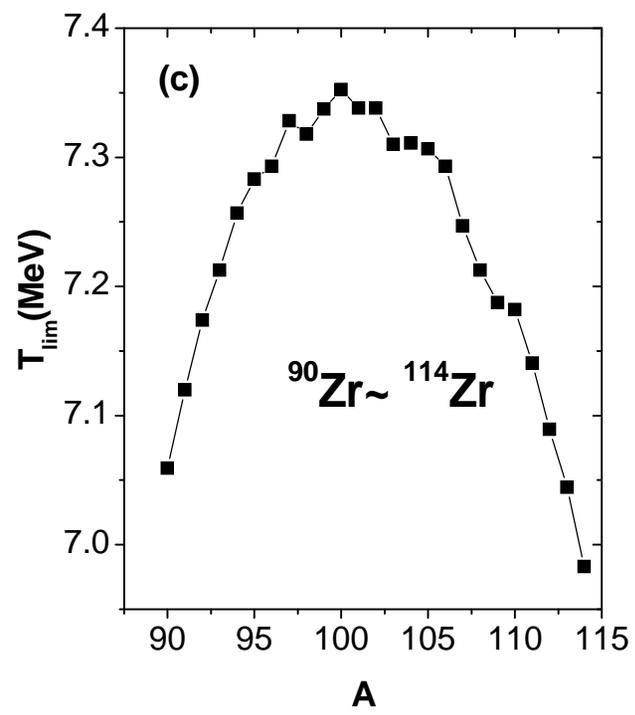
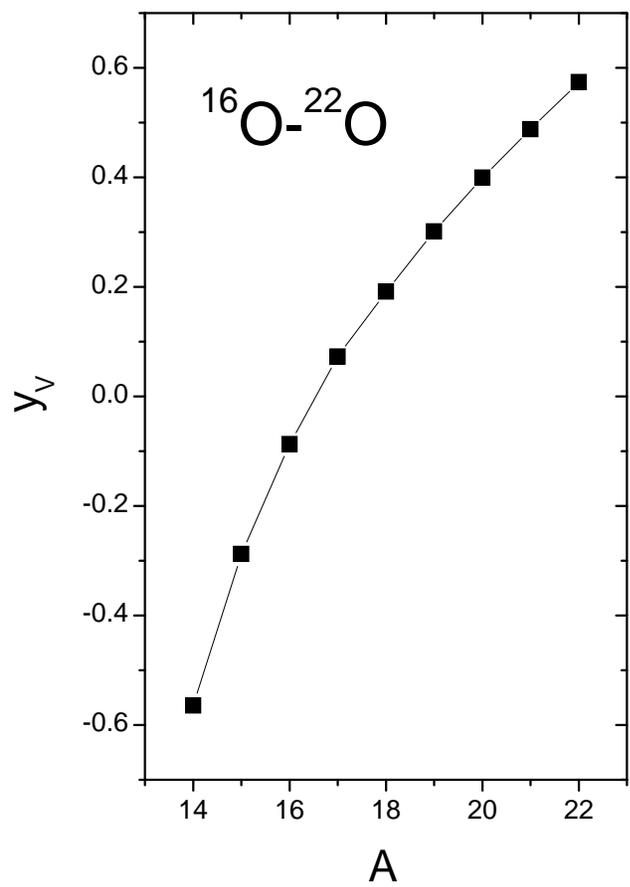
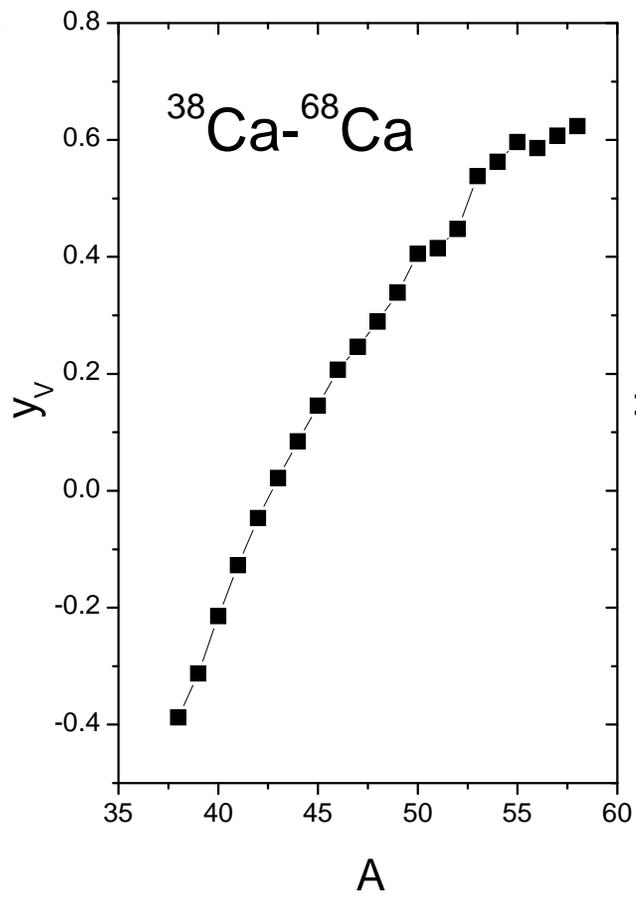
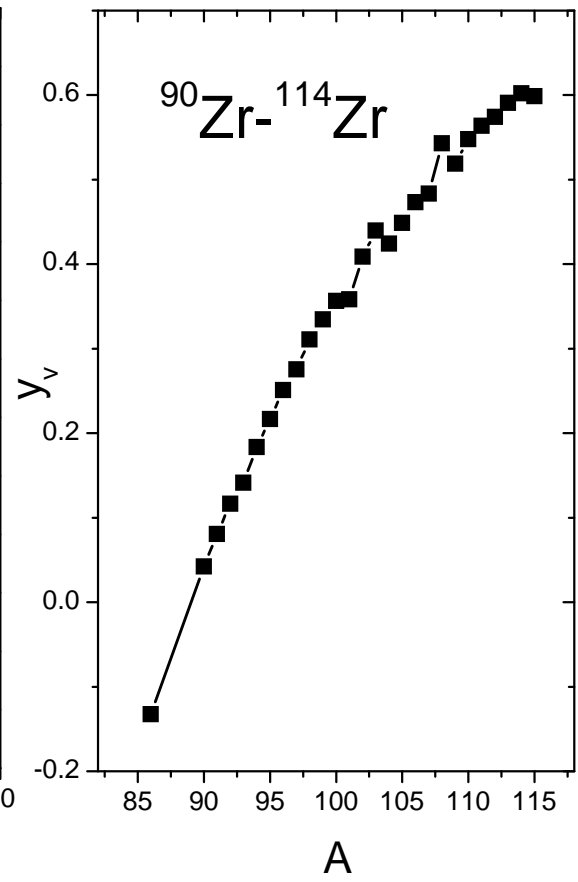
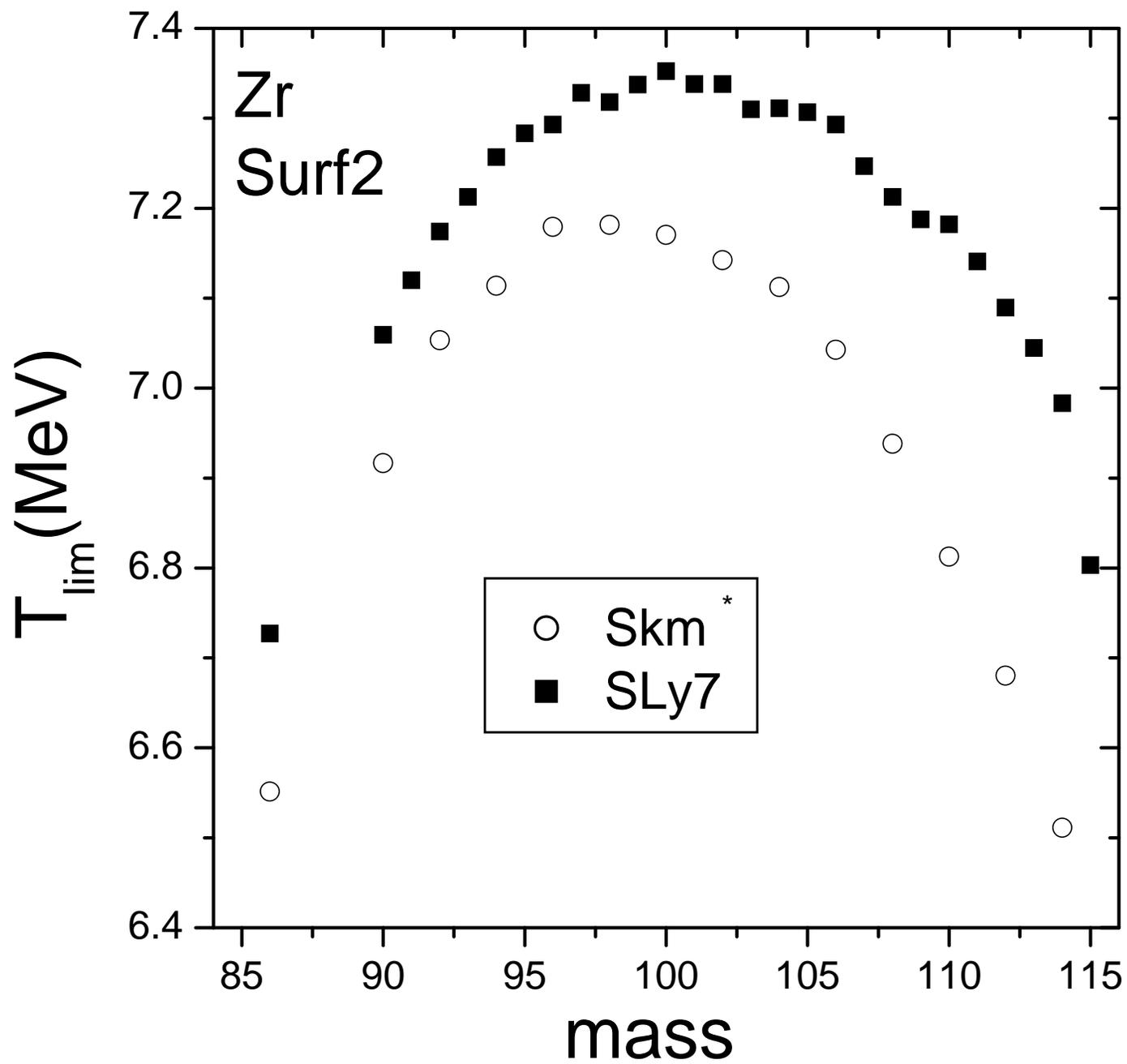
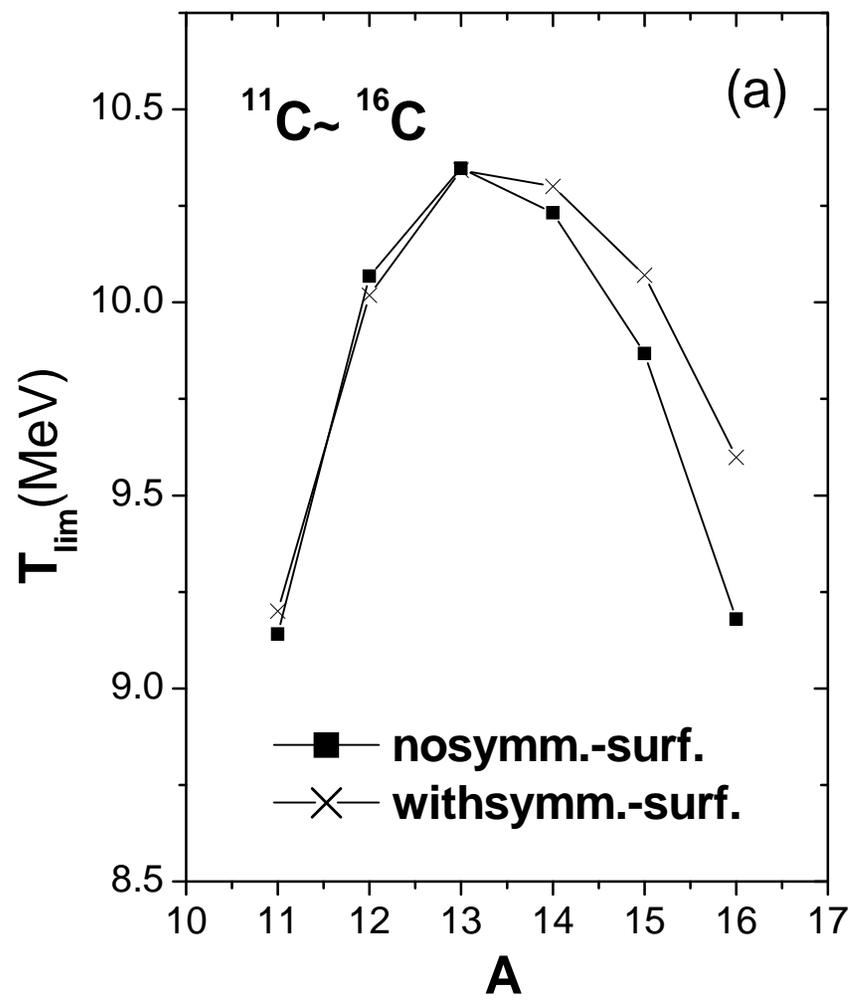
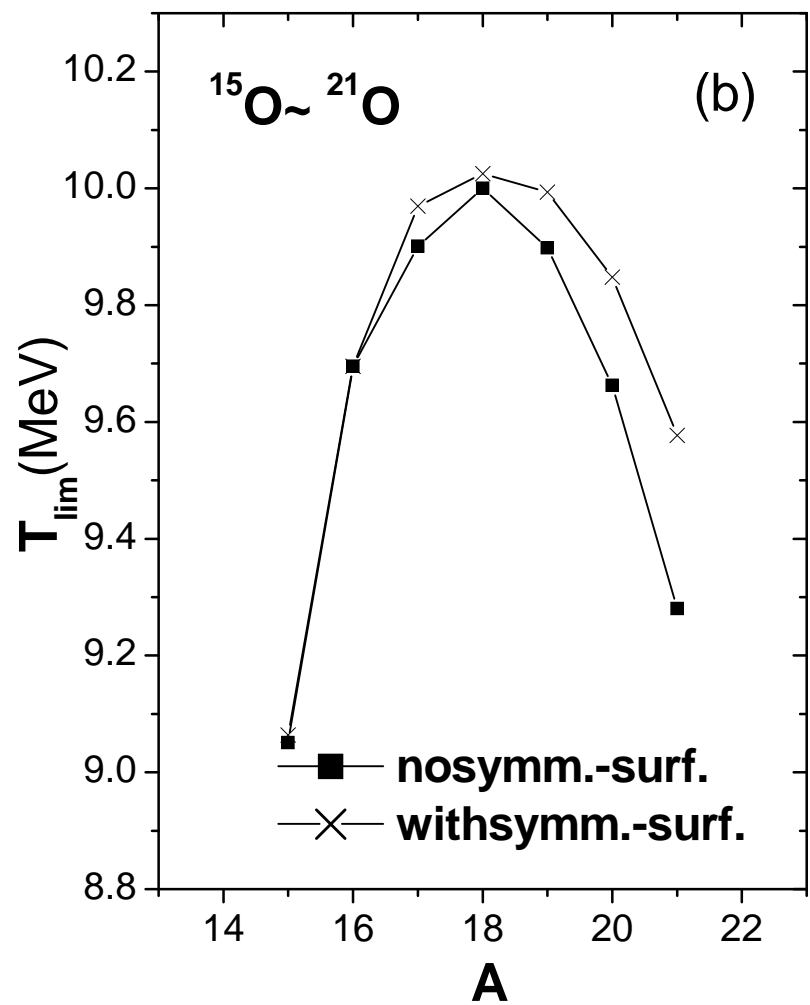
\end{document}